\documentclass[twocolumn,prl]{revtex4}
\usepackage{graphicx}
\usepackage{dcolumn}% Align table columns on decimal point
\usepackage{bm}% bold math
\usepackage{amsmath}
\usepackage{times}
\usepackage{epstopdf}
\usepackage{color}
\usepackage[breaklinks=true,colorlinks,citecolor=blue,linkcolor=blue,urlcolor=blue]{hyperref}

\makeatletter

\newcommand{\Rmnum}[1]{\expandafter\@slowromancap\romannumeral #1@}
\makeatother
\graphicspath{{images/}}
\begin{document}

\title{Antiferromagnetic rare region effect in Pr$_{0.5}$Ca$_{0.5}$MnO$_3$}

\author{Vinay Kumar Shukla}
\email{vkshukla@iitk.ac.in}
\affiliation{Department of Physics, Indian Institute of Technology, Kanpur, 208016, India}
\author{Soumik Mukhopadhyay}
\email{soumikm@iitk.ac.in}
\affiliation{Department of Physics, Indian Institute of Technology, Kanpur, 208016, India}

\begin{abstract}
We present evidence of coexistence of electron paramagnetic resonance signal and anti-ferromagnetic resonance signals above the antiferromagnetic (AFM) transition ($T_{N}$) in $Pr_{0.5}Ca_{0.5}MnO_{3}$. We identify the latter with AFM rare regions within the `Griffiths-like' phase scenario with the associated temperature scale T$^\ast$ extending above room temperature.
\end{abstract}

\maketitle

After going through several decades of intense experimental and theoretical scrutiny, there is a general acceptance of certain features which are near universal in mixed-valent manganites~\cite{Dagotto, Tokura0} with far reaching implications for complex systems in general~\cite{Dagotto1}: 1) Prevalence of the nanoscale phase separation; 2) Existence of a higher temperature scale T$^\ast$, where short range `clustering' sets in, in addition to the long range ordering temperature. However, the underlying physical model that describes such a scenario and consequently the origin of colossal magnetoresistance (CMR) in manganites is still intensely debated~\cite{Souza, Dagotto}. At present there are primarily two hypotheses: 1) The existence of Griffiths-like phase with percolation type metal-insulator (MI) transition~\cite{Salamon, Krivoruchko1, Krivoruchko2}; 2) Alternatively, that the Griffiths phase (GP) in itself is insufficient to cause CMR, and the formation of ferromagnetic polarons just above the long range ordering temperature needs to be taken into account~\cite{Souza, Jiyu, Ivanshin}. However, for low band width systems near half doping, the debate is mainly centered around the following competing pictures of charge ordering~\cite{Tokura}: 1) CE type charge and magnetic ordering, originally proposed by Goodenough~\cite{Goodenough}; 2) Zener polaron (ZP) ordering with antiferromagnetic interaction between strong ferromagnetically coupled dimers~\cite{Aladine}; 3) Coexistence of both CE type and ZP type ordering~\cite{Khomskii} or between correlated and uncorrelated polarons~\cite{Jooss}. The last two frameworks have the additional advantage of being able to explain away the emergence of ferroelectricity in charge-ordered manganites~\cite{Yamauchi, Khomskii, Shukla1, Shukla2}.

The existence of a phase similar to the Griffiths phase in manganites is confirmed by the following experimental signatures: 1) Presence of a weak, usually ferromagnetic resonance (FMR) signal above T$_C$ against the background of an electron paramagnetic resonance signal~\cite{Deisenhofer1}; 2) Deviation from typical behavior predicted by the standard theory of second order phase transitions so far as the critical indices in the temperature dependence of the magnetic susceptibility in the paramagnetic region is concerned~\cite{Krivoruchko1}, 3) The downturn in the inverse susceptibility as compared to the paramagnetic Curie background is suppressed by increasing the magnetic field due to the increased magnetization of the paramagnetic matrix enclosing the rare region.

Quenched disorder is a prerequisite for the formation of the GP, although the physical picture of short-range-correlated disorder creating large scale spin and charge inhomogeneities in manganites is only applicable to a narrow window at low doping and should be absent near half doping~\cite{Bouzerar}. However, recently there have been a few experimental studies which claim existence of GP in half doped systems with intermediate band width~\cite{Pramanik1, Pramanik2}. It has been predicted that the coexistence of two competing phases separated by a first order transition enhances the formation of a `Griffiths-like' clustered phase below a characteristic temperature T$^\ast$~\cite{Burgy1}. The coexistence of ferromagnetic metallic (FM) and antiferromagnetic (AFM) insulating phase is ubiquitous in manganites~\cite{soumik1, soumik2} with AFM phase dominating close to half doping~\cite{Brink}. Given such a scenario, it is surprising that occurrence of AFM rare regions in manganites has not been reported so far. In this letter, we provide experimental evidence of AFM rare regions in a narrow band width manganite at half doping, namely, Pr$_{0.5}$Ca$_{0.5}$MnO$_{3}$.

Pr$_{1-x}$Ca$_{x}$MnO$_{3}$ ($0.3<x<0.5$) in bulk form shows transition from paramagnetic (PM) to AFM phase at low temperature (T$_N$) intermediated by the onset of charge ordering at a higher temperature ($T_{CO}>T_{N}$). Poly-crystalline Pr$_{1-x}$Ca$_{x}$MnO$_{3}$ (PCMO) samples with $x=0.5,0.45,0.4,0.33$ were prepared by the standard method described elsewhere~\cite{Shukla1}. The structural characterization of all the samples were done by x-ray diffraction $\theta$- 2$\theta$ scans at room temperature using PANalytical X'pert diffractometer with $Cu-K_\alpha$ radiation having wavelength of 1.54 \AA. The Rietveld refinement analysis done by using full prof suite reveals that the room temperature phase of all the samples has orthorhombic structure having \textit{Pbnm} space group symmetry. The microstructure, crystallite size and its distribution were studied by field emission scanning electron microscope (FE-SEM, Jeol, JSM-7100F). The chemical composition of all the samples were confirmed by energy dispersive spectroscopy (EDS) and the x-ray photoemission spectroscopy using PHI 5000 Versa Prob II, FEI Inc. See Ref.~\cite{SM} for details of structural and chemical characterization. The magnetic measurements were carried out in a Quantum Design PPMS. The temperature dependent EPR spectroscopy was done done using Bruker EPR EMX spectrometer in the X-band.
\begin{figure}
\includegraphics[width=8.5 cm]{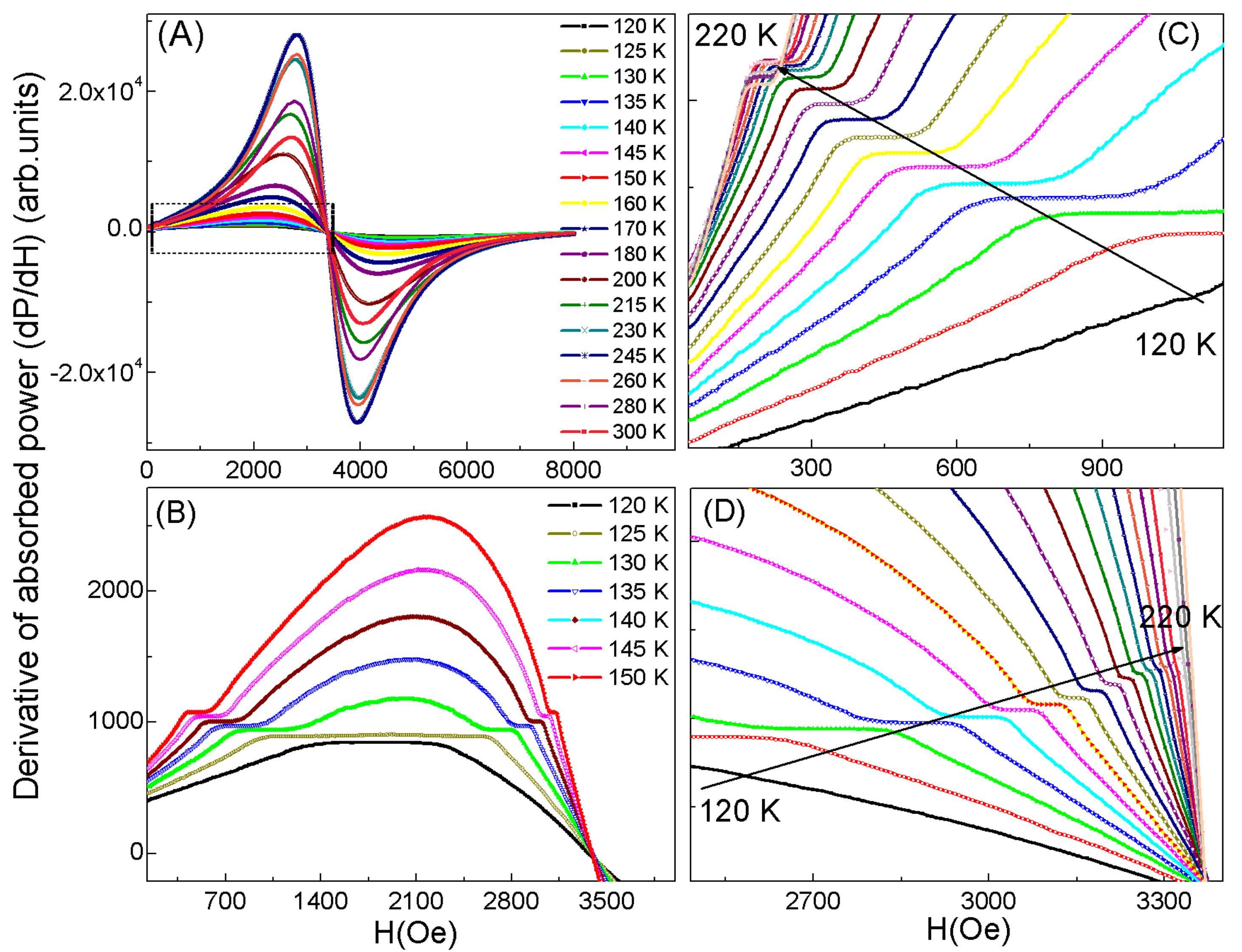}
\caption{(A) ESR signals shown for PCMO(x=0.5, bulk) from 120 K to 300 K, (B) The area highlighted in A) is zoomed in to show the additional resonance peaks emerging from the main paramagnetic signal. (C) Low field (LF) resonance field shifts towards lower magnetic field value with increasing the temperature from 120 K to 220 K (indicated by solid black arrow), and (D) Shift of high field (HF) resonance position towards higher H value with increasing the temperature from 120 K to 220 K is shown. Above 220 K, HF signals could not be distinguished anymore.}\label{fig:esr1}
\end{figure}
\begin{figure}
\includegraphics[width=8.5 cm]{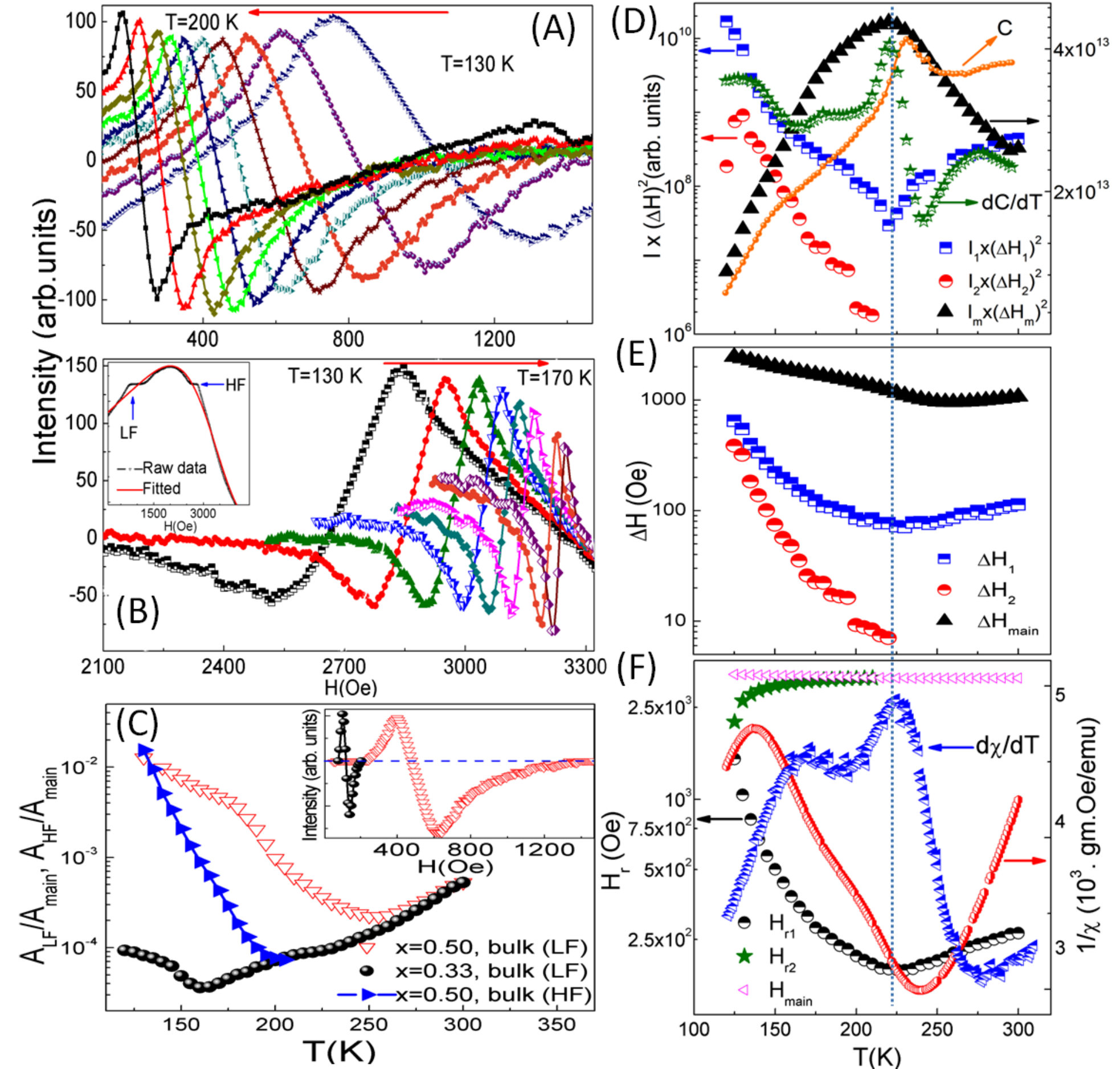}
\caption{(A) and (B) represents the temperature evolution of LF and HF signals after extraction from the main resonance peak. LF and HF peaks are obtained by subtracting the main Lorentzian signal from the raw data as shown in inset of (B). (C) ratio of area under the LF (A$_{LF}$) and HF (A$_{HF}$) signals normalized with respect to the corresponding main resonance signal (A$_{main}$) is shown for PCMO (x=0.5, bulk) and PCMO (x=0.33, bulk). The area under the LF peak of PCMO (x=0.33, bulk) is significantly reduced as compared to that of PCMO (x=0.5, bulk) (shown in inset of (C)). (D) Temperature dependence of the product of intensity and square of linewidth for LF ($I_1\times(\Delta H_1)^2$) and HF ($I_2 \times (\Delta H_2)^2$) peaks along with main resonance ($I_{main} \times (\Delta H_{main})^2$) is shown. The temperature dependence of heat capacity and its derivative (calculated from the standard data in Ref.~\cite{Hardy}) are also plotted. (E) The line-widths of LF ($\Delta H_1$) and HF ($\Delta H_2$) signals are compared with the main ($\Delta H_{main}$). (F) The corresponding resonance fields of LF ($H_{r1}$) and HF ($H_{r2}$) signals along with the main ($H_{main}$) resonance are shown. The inverse of dc magnetic susceptibility ($1/\chi$) along with its first derivative (d$\chi$/dT) for PCMO (x=0.5, bulk) are also plotted in (F). The vertical dotted blue line represents the temperature $220$ K.}\label{fig:esr2}
\end{figure}

\begin{figure}
\includegraphics[width=8.5 cm]{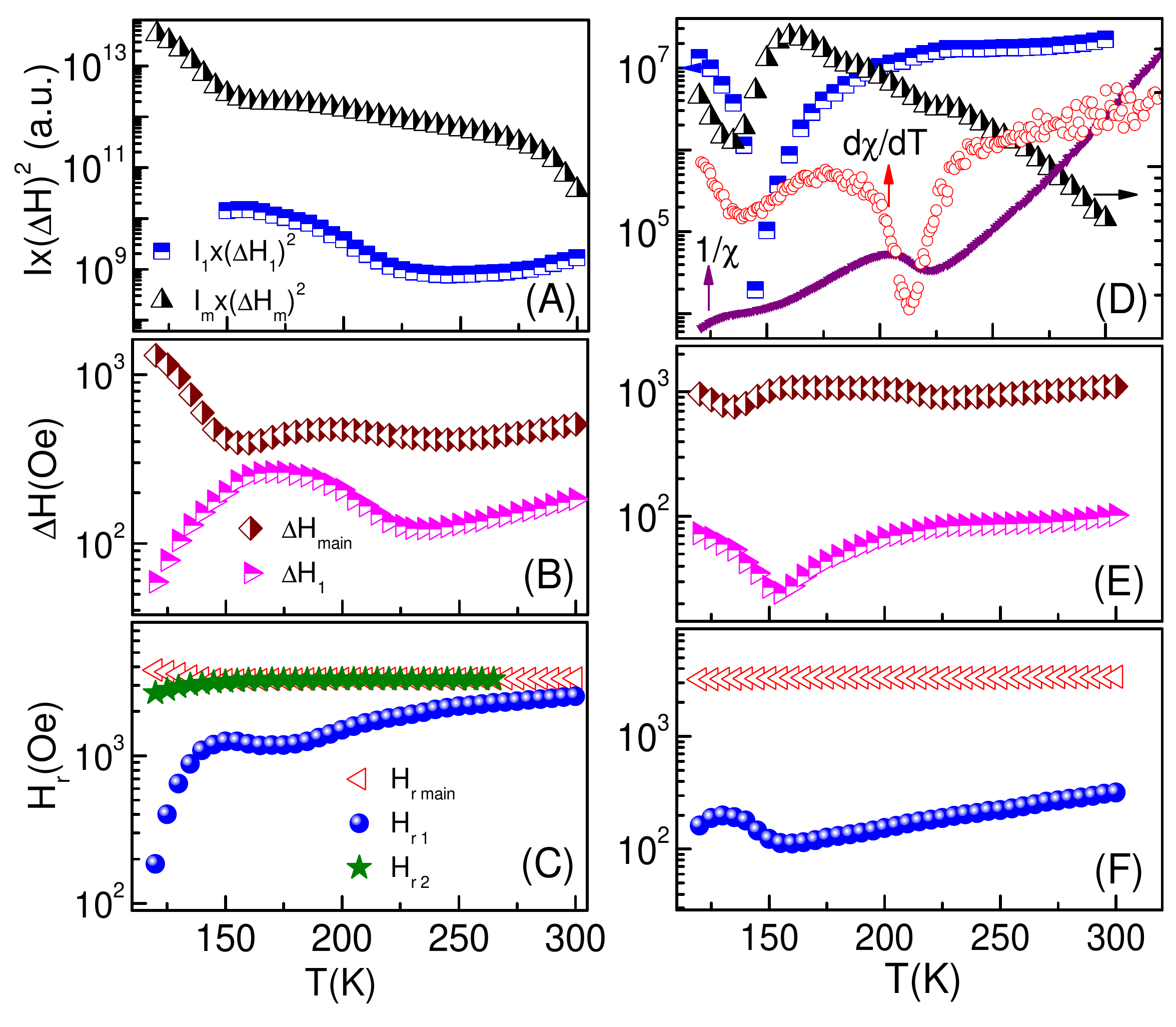}
\caption{(A) The product of the ESR intensity and square of linewidth (I x ($\Delta H)^2$) for LF and main resonance peak for PCMO (x=0.40, bulk). (B), (C) Corresponding linewidths and resonance fields, respectively. (D), (E), and (F) Plots of the same for PCMO (x=0.33, bulk).}\label{fig:esr3}
\end{figure}

The differential EPR signals were recorded at different temperatures from 120 K up to room temperature for the PCMO samples. The samples were exposed to microwave radiation at constant frequency of 9.46 GHz (X-band) and external magnetic field was varied from 0 to 8000 Gauss. The power (P) absorbed by the sample from the transverse magnetic microwave field is captured in the form of its first derivative (dP/dH) by the standard lock-in technique~\cite{Poole, John}. Fig.~\ref{fig:esr1}A shows the EPR signals of $Pr_{0.5}Ca_{0.5}MnO_{3}$ at some representative temperatures between 120-300 K. In general, the line-shape is symmetric Lorentzian. The origin of EPR signals in these systems is generally attributed to the combined effect of $Mn^{3+}$ and $Mn^{4+}$ states (which are coupled through double exchange interaction) and the lattice~\cite{Lofland, Causa, Shengelaya1, Shengelaya2}. Strikingly, we observe the appearance of a pair of additional resonance peaks as shown in Fig.~\ref{fig:esr1}B. It is clear that the low field (LF) and high field (HF) resonance positions approach each other with lowering of temperature (Fig.~\ref{fig:esr1}C, D). In order to understand the nature of these resonance peaks and to accurately calculate corresponding intensities, linewidth and resonance fields, we have fitted the ESR signals by the following equation~\cite{Ivanshin1, Liu1}-
\begin{equation}\label{eq:1}
\frac{dP}{dH}\propto \frac{d}{dH}\left(\frac{\Delta H}{(H-H_r)^2 +\Delta H^2}+ \frac{\Delta H}{(H+H_r)^2 +\Delta H^2}\right)
\end{equation}
where $H_r$ is the resonance field and $\Delta H$ is the linewidth. The resonance peaks are extracted by subtracting the main PM resonance signal described by the Lorentzian in equation~\ref{eq:1} from the raw data. As a result, we obtain two sets of peaks, namely, LF and HF peaks on either side of the positive half maximum (around H$\sim$2180 Oe) of the main resonance peak (Fig.~\ref{fig:esr2}A,B). The intensity, linewidth and resonance fields of LF and HF signals have been calculated by integrating the LF and HF spectra and fitting the integrated signals with Lorentzian line shape function given in the equation~\ref{eq:1}. The product of intensities and linewidth squared ($I_{max}\times{\Delta H}^2$) for the LF and HF spectra and main resonance peak are shown in Fig.~\ref{fig:esr2}D. While the $I_{max}\times{\Delta H}^2$ for the LF resonance shows a sharp anomaly near $220$K, the signal for HF resonance is not observable beyond $220$ K with the corresponding resonance field merging with the paramagnetic backbone. Curiously, the intensity for main resonance shows a maximum again at the same temperature. The temperature dependence of line-width, resonance fields for the three signals along with the temperature derivative of dc susceptibility ($\frac{d\chi}{dT}$) in the same temperature range are plotted in Fig.~\ref{fig:esr2}E, F. Strikingly, the maximum in $\frac{d\chi}{dT}$, too, appears at $220$K (Fig.~\ref{fig:esr2}F). We plot the temperature derivative of total heat capacity ($C$), calculated from the standard literature data~\cite{Hardy}, in Fig.~\ref{fig:esr2}D where the global maximum in $\frac{dC}{dT}$ again appears around $220$K, serving as further evidence of a thermodynamic phase transition. We observe similar resonance signals in the EPR spectra for other PCMO samples (x=0.45,0.4,0.33) as well. However, the signals are weaker compared to that observed in Pr$_{0.5}$Ca$_{0.5}$MnO$_3$. For PCMO (x=0.4, bulk), we can precisely determine the temperature dependence of $\Delta$H and intensities only for the LF signal although the H$_r$ values of the LF as well as the HF peaks can be estimated. The temperature dependence of $I_{max}\times{\Delta H}^2$ and H$_r$ for the LF peak of PCMO (x=0.40, bulk) and PCMO (x=0.33, bulk) as well as the anomalies observed in the temperature dependence of linewidth are similar to that of the half doped PCMO (Fig.~\ref{fig:esr3}). The striking correlation between the temperature dependence of H$_r$, $\Delta$H and I$_{max}$ on one hand and the macroscopic dc susceptibility data on the other observed for the half doped system (Fig.~\ref{fig:esr2}), is missing away from half doping (Fig.~\ref{fig:esr3}E). Moreover, for $x=0.33$, the HF signal is completely suppressed.

\begin{figure}
\includegraphics[width=8.5 cm]{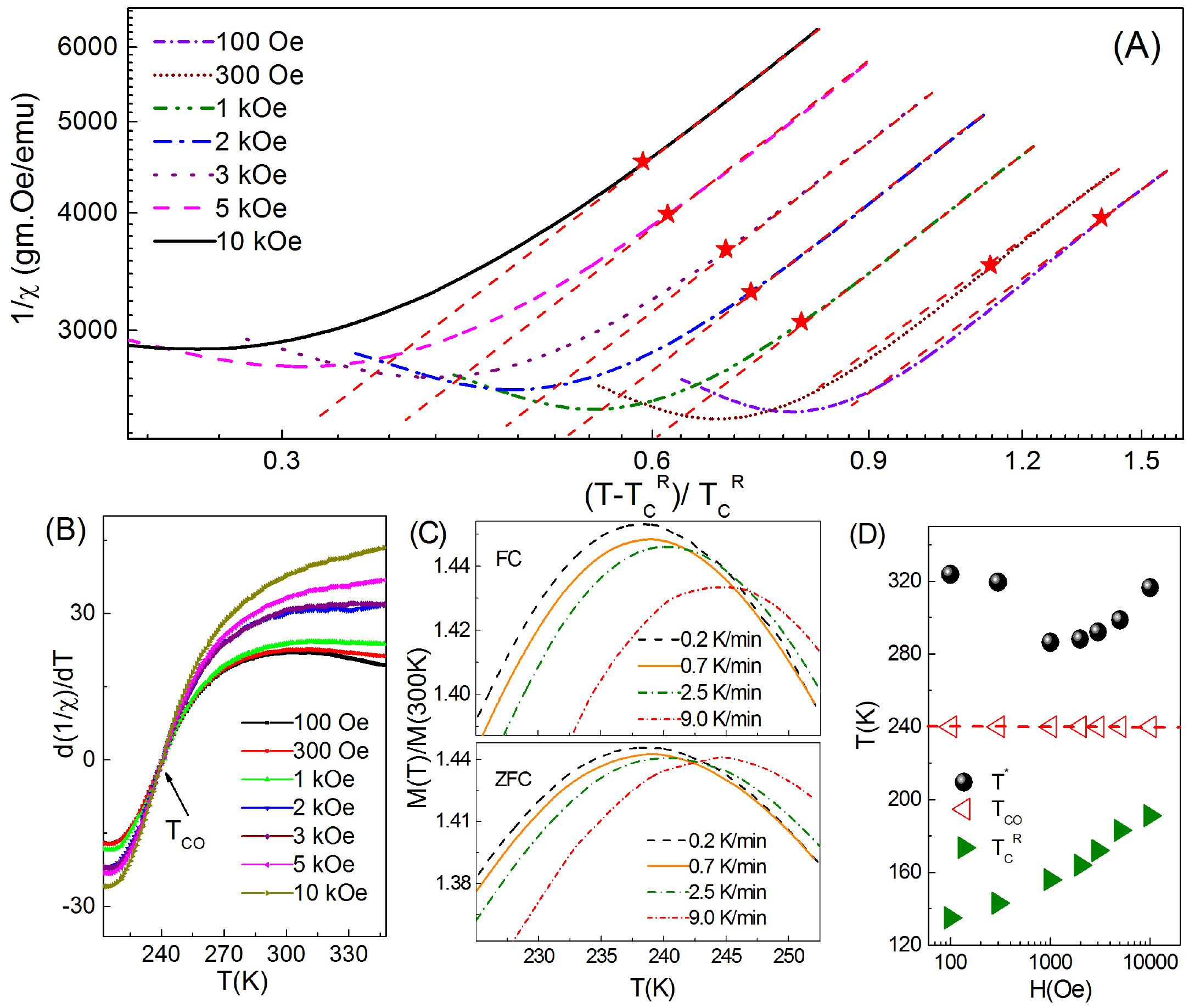}
\caption{(A) The log-log plots of field cooled (FC) inverse dc magnetic susceptibility (1/$\chi$) plotted against reduced temperature (T-$T_C^R$)/ $T_C^R$ at different applied magnetic field ranging from 100 Oe to 10 kOe. The Curie-Weiss fits are shown by dashed red line and star symbols represents the location of corresponding $T^\ast$ for different magnetic fields. (B) Temperature dependence of the first derivative of inverse dc susceptibility ($d(1/\chi)/dT$) at different applied magnetic fields. The charge ordering temperature $T_{CO}$ is indicated by the arrow.(C) The temperature dependence of reduced magnetization (M(T)/M(300K)) (FC and ZFC) at different temperature sweep rates ranging from 0.2 K/min to 9.0. K/min. (D) Variation of fitting parameters $T^\ast$, $T_{CO}$, and $T_C^R$  as a function of applied magnetic field.}\label{fig:gpanalysis}
\end{figure}
\begin{figure}
\includegraphics[width=8.5 cm]{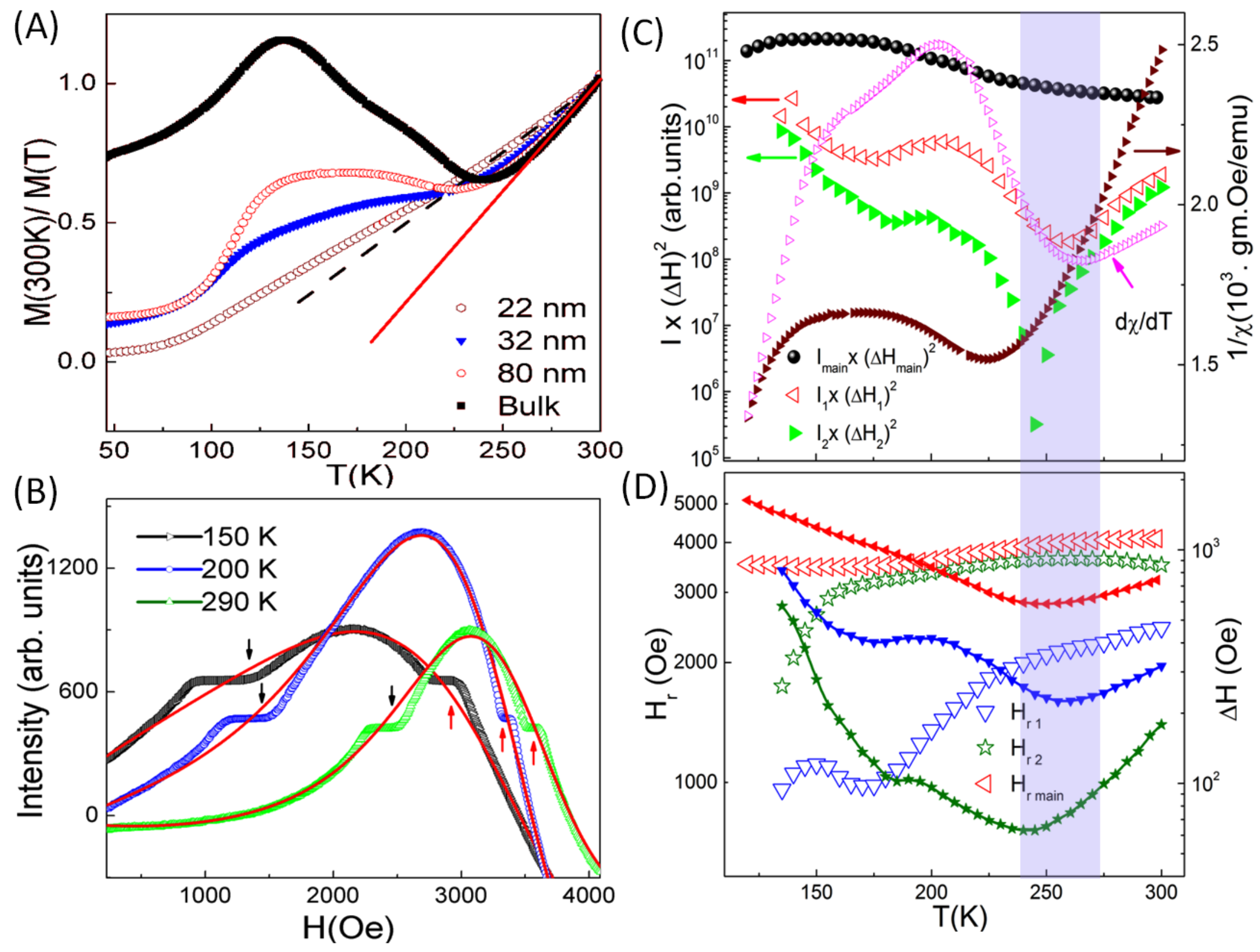}
\caption{(A) The inverse field cooled (FC) magnetization normalized with respect to its value at 300 K is shown for different crystallite sizes of PCMO (x=0.5) at H=1 kOe. The solid red and dashed black lines represents the fits to curie law. (B) The ESR signals of PCMO (x=0.5, 80 nm) at some representative temperatures along with LF and HF peaks marked by black down and red up arrows respectively. The solid red lines represents the corresponding fits to ESR data. (C) Temperature dependence of the product of intensity and square of line-width for LF ($I_1 \times (\Delta H_1)^2$) and HF ($I_2 \times (\Delta H_2)^2$) signals along with that of the main resonance ($I_{main} \times (\Delta H_{main})^2$) are plotted. The inverse of dc magnetic susceptibility ($1/\chi$) along with its first derivative (d$\chi$/dT) for PCMO (x=0.5, 80 nm) are also plotted over the same temperature scale. The resonance fields of LF ($H_{r1}$), HF ($H_{r2}$) and main signals ($H_{main}$) along with corresponding linewidths (filled symbols) are shown in (D).}\label{fig:gpnano}
\end{figure}
The coexistence of LF and HF signal in the intermediate temperature regime with the former extending up to the room temperature suggests existence of a complex magnetic phase unlike ever reported before in manganites. To the best of our knowledge, in existing reports on manganites, the number of additional resonance signal observed other than the PM signal is restricted to one and that, too, for low doped systems~\cite{Deisenhofer1}. We shall take up the issue of emergence of these signals in a more elaborate manner shortly hereafter. Let us, for the moment, turn our attention to the dc susceptibility data above $T_{CO}$ (Fig.~\ref{fig:gpanalysis}A). Generally one expects a downturn in the temperature dependence of inverse susceptibility due to the growth of ferromagnetic clusters with lowering of temperature in the Griffiths phase. However we observe that the downturn above $T_{CO}$ is only marginal which is followed by a sharp upturn just below $T_{CO}$. The marginal downturn above $T_{CO}$ is completely suppressed at slightly higher magnetic field eventually leading to an upward deviation over the paramagnetic Curie background. Prima facie, this suggests existence of AFM rare regions above $T_{CO}$ as the AFM susceptibility for the rare regions should be less than or comparable to the paramagnetic susceptibility such that when they add up, the downturn is not so pronounced as observed in case of FM Griffiths phase. Moreover, it is difficult to envisage a finite FM rare region, since half doped PCMO is known to exhibit electronic phase separation with a spatial distribution of hole concentration only, without introducing any FM phase~\cite{Brink, Tomioka}.

The identification of the LF and HF signals with AFM cluster phase is supported by the following observations: 1) The coexistence of two symmetrically placed resonances typical of AFM resonance spectra where application of external field increases the effective field of one component while decreasing the same for the other. In the present case, the two resonance fields have opposite temperature dependence (Fig.~\ref{fig:esr1}, Fig.~\ref{fig:esr2}F); 2) For the LF signal in EPR spectra, the resonance field (H$_r$) decreases marginally with lowering of temperature with concomitant sharp reduction in LF signal intensity down to $220$ K. The corresponding line-width for the LF signal shows a minimum at the same temperature suggesting exchange narrowing effect. 3) Although it is rare to observe AFM resonance signal in the X band, for small clusters, the anisotropy field $H_A$ could be low enough to push the AFM resonance towards the X-band. Indeed, multiple resonance signals, excluding the paramagnetic one in the X-band, have been associated with short range AFM correlations~\cite{Gareth, Yang}. Below $220$K, a sharp increase in the LF and HF signal intensity is observed (Fig.~\ref{fig:esr2}D), which suggests some canting instability in the AFM phase. The effect of canting instability is stronger at low temperature as supported by the gradual reduction in the difference between $H_r$ values in the two spectra with lowering of temperature accompanied by an increase in the linewidth. If we look at the ratio of area under the LF signal to that of the main signal (Fig.~\ref{fig:esr2}C), the ratio first slowly decreases with lowering of temperature followed by a significant upturn below T$_{CO}$, suggesting the growing contribution of the LF signal at the expense of the main PM signal with lowering of temperature in this temperature regime. The HF signal, too, grows at the expense of PM signal, eventually catching up with the LF signal at low temperature. Away from half doping, however, the contribution of LF signal vis$\mathit{\mbox{-}}$$\grave{a}$$\mathit{\mbox{-}}$vis PM signal is considerably reduced (Fig.~\ref{fig:esr2}C). A rough estimation by comparing LF/HF signals with PM signal gives the fraction of spins contributing to LF signal for half doped PCMO to be $\leq$ 1.3\%, whereas the fraction of spins contributing to HF signals is $\leq$ 1.6\%. Away from half doping, the corresponding fraction of LF signals to the main signals are estimated to be $\leq$ 0.15\% for $x=0.4$ and $\leq$ 0.05\% for $x=0.33$, respectively. This is consistent with our observation that rare regions are not dominant away from half doping.

In order to check whether the macroscopic susceptibility is influenced by cluster effects so clearly observed in the EPR spectra, one needs to correlate the temperature dependence of various parameters obtained from the resonance signals with the temperature dependence of macroscopic inverse susceptibility. The marginal downturn in the temperature dependence of inverse susceptibility above T$_{CO}$ is suppressed at higher magnetic field due to the growth of PM signal. Below $T_{CO}$, the strengthening of the AFM contribution at the expense of the PM signal leads to the upturn in inverse susceptibility. That there is a strengthening of AFM cluster phase is further supported by the fact that the linewidth decreases as temperature is lowered towards $220$K. Interestingly, the first order derivative of inverse susceptibility with respect to the temperature is completely insensitive to the application of magnetic field only at T$_{CO}$ whereas above T$_{CO}$, $\frac{d(1/\chi)}{dT}$ shows a local maximum at low field reaffirming the downturn of inverse susceptibility which is progressively suppressed at higher magnetic field (Fig.~\ref{fig:gpanalysis}B). Moreover, as discussed earlier, the global maxima in the temperature dependence of $\frac{d(1/\chi)}{dT}$ and $\frac{dC}{dT}$ coincide exactly with the sharp anomaly in the LF signal intensity, the minima in $\Delta H$ and the maximum and minimum in $H_{r1}$ and $H_{r2}$ respectively suggesting strong influence of the AFM rare region on the macroscopic susceptibility and a phase transition at $220$K associated with the rare region. Although the LF and HF signals lie below the PM signal, the LF and HF resonance fields should ideally be compared with the PM signal above T$^\ast$ which is outside the temperature range for EPR measurements. Since the HF signal is very close to the PM signal in the same temperature range, it is possible that resonance field for HF signal might actually be higher than that for the PM signal above T$^\ast$. There could be an alternative scenario as the total number of AFMR signals might be more than two as observed for orthorhombic symmetry before~\cite{Yamauchi1}. In that case the two observable AFMR signals can lie below the resonance field for the paramagnetic signal.

We further analyze the data by fitting the temperature dependence of inverse dc magnetic susceptibility using the relation~\cite{Griffiths, Bray1, Castro, Bray2, Jiang} 
\begin{equation}
\chi\propto \frac{1}{(T-T_C^R)^{1-\lambda}}
\end{equation}
where $\lambda$ lies between 0 and 1. The temperature below which there is deviation from the Curie-Weiss fit determines the temperature scale $T^\ast$ for cluster formation as shown in the log-log plots of 1/$\chi$ against the reduced temperature (T-${T_C}^R$)/ ${T_C}^R$ for different magnetic fields in Fig.~\ref{fig:gpanalysis}A. We identify the value of ${T_C}^R$ for which the slope (1-$\lambda$) of fitted data above T$^\ast$ is unity (i.e. $\lambda=0$). Once we obtain the value of ${T_C}^R$, the value of $\lambda$ can be extracted from the slope of low temperature side of the data below T$^\ast$. However we do not find any distinct power law behavior in that regime. The extracted parameters ${T_C}^R$, T$_{CO}$ and $T^\ast$ are plotted with respect to the magnetic field in Fig.~\ref{fig:gpanalysis}D. While $T_{CO}$ is independent of the magnetic field, the value of T$^\ast$ initially decreases at low magnetic field before increasing at higher H.

To distinguish between conventional second order magnetic transition described by the polaron picture and the Griffith's like scenario, we study the time relaxation of zero field cooled (ZFC) and field cooled (FC) magnetization, since Griffith's like state is prone to exhibiting out-of-equilibrium features due to anomalously slow relaxation of magnetization~\cite{Vazquez}. The temperature dependence of dc magnetization (FC and ZFC) is shown for different sweep rates ranging from 0.2 K/min to 9.0. K/min in Fig.~\ref{fig:gpanalysis}C. With increasing the temperature sweep rate from 0.2 K/min to 9.0 K/min, the magnetic anomaly associated with charge ordering shifts towards higher temperature for both FC and ZFC magnetization. Such sensitivity of FC magnetization to the temperature sweep rate is not expected for a conventional second order magnetic transition. Interestingly, for FM Griffiths phase, we should expect the anomaly to shift to lower temperature with increasing sweep rate~\cite{Krivoruchko3, Krivoruchko4}, exactly opposite to our observation. Although, a theoretical treatment is lacking, we emphasize that such a response could be due to the AFM nature of the rare region.

In the end, an important point remains to be addressed. If the LF and HF signals are attributed to the AFM rare region, then one should expect a common temperature range for both, which is clearly not the case for the bulk half-doped PCMO. One possibility is that as the HF signal shifts towards higher resonance field with increasing temperature, it eventually approaches the main resonance asymptotically, thus making it impossible to distinguish between the two. Fig.~\ref{fig:gpnano}A shows the inverse susceptibility data for poly-crystalline PCMO with different average grain size along with the bulk. Except for the lowest grain size, the magnetic anomaly related to charge ordering survives in all other samples. And indeed, as the average grain size is lowered to 80 nm, we find that both LF and HF signals extend at least up to room temperature (Fig.~\ref{fig:gpnano}B,C,D). On further reduction of grain size, the additional resonance signals disappear altogether (not shown in Figure). The variation of different parameters extracted from the LF and HF signal is similar to the bulk poly-crystalline sample although the striking correlation with $\frac{d\chi}{dT}$ is missing (Fig.~\ref{fig:gpnano}C,D). The anomalies in $\Delta H$, $I_{max}\times \Delta H^2$, etc are instead confined within a broad temperature region around the minima in $\frac{d\chi}{dT}$. A comparison of the LF/HF signals and PM signals shows
that the fraction of spins contributing to the LF and HF resonances are $\leq1.7\%$ and $\leq1.65\%$, respectively, which are slightly higher than the corresponding bulk sample. The mismatch of the anomalies in the macroscopic susceptibility and the LF/HF signal parameters can be attributed to the increased FM correlation in the main signal due to reduction in grain size~\cite{Vinay} as well as the distribution of the grain size in the nanocrystalline sample.

To conclude, we present direct experimental evidence of AFM rare region effects above T$_{N}$ in half doped narrow band width PCMO with the associated temperature scale $T^\ast$ extending above room temperature. In a nutshell, the various findings are as follows: 1) Observation of a pair of resonance signals (extending at least up to room temperature) other than the main paramagnetic resonance; 2) Marginal downward deviation at low magnetic field from the Curie background in the inverse susceptibility, which is suppressed at higher magnetic field with $T^\ast$ showing strong non-monotonic magnetic field dependence; 3) Slow time relaxation in the field cooled magnetization even far above the AFM ordering temperature; 4) Although we fail to ascertain any power law behaviour well below $T^\ast$, there is a strong correlation between the temperature evolution of the independently measured AFM resonance signals and the macroscopic susceptibility. It seems highly probable that the physics of AFM rare regions within the Griffiths phase scenario is applicable to other low and intermediate bandwidth manganites near half doping as well, something which needs to be verified experimentally in future.

\end{document}